\documentclass{article}

\usepackage{microtype}
\usepackage{graphicx}
\usepackage{subfigure}
\usepackage{booktabs} 
\usepackage{caption}
\usepackage{hyperref}
\usepackage{bbm}
\usepackage{multirow}
\usepackage{makecell}

\usepackage[toc,page]{appendix}
\usepackage[accepted]{icml2024}

\usepackage{array}
\usepackage{xspace}

\usepackage{amsmath}
\usepackage{amssymb}
\usepackage{mathtools}
\usepackage{amsthm}

\usepackage{stfloats,eucal}

\usepackage{colortbl}
\usepackage{xcolor}
\usepackage{color}
\definecolor{lm_purple_low}{RGB}{240,240,248}
\def\ie{{\it i.e.}\xspace}
\usepackage[capitalize,noabbrev]{cleveref}

\def\pM{{\mathcal{M}}}
\def\pS{{\mathcal{S}}}
\def\pP{{\mathcal{P}}}
\def\pX{{\mathbf{X}}}
\def\pR{{\mathbf{R}}}
\def\br{{\mathbf r}}
\def\pT{{\mathbf{T}}}
\def\pA{{\mathcal{A}}}
\def\pD{{\mathcal{D}}}
\def\prm{{\pR_\pM}}
\def\prs{{\pR_\pS}}
\def\pxm{{\pX_\pM}}

\def\dX{{\mathbf X}}
\def\bx{{\mathbf x}}
\def\rn{{\pR_\delta}}
\def\tn{{\pX_\delta}}
\def\rnh{{\hat{\pR}_\delta}}
\def\tnh{{\hat{\pX}_\delta}}

\def\Ro{{\pR^{(t-1)}}}
\def\Roo{{\pR^{(0)}}}
\def\Ri{{\pR^{(t)}}}
\def\Roh{{\hat{\pR}^{(t-1)}}}
\def\Rooh{{\hat{\pR}^{(0)}}}

\def\Rlo{{\pR^{(l-1)}}}
\def\Rli{{\pR^{(l)}}}
\def\Xo{{\pX^{(t-1)}}}
\def\Xoo{{\pX^{(0)}}}
\def\Xi{{\pX^{(t)}}}
\def\Xoh{{\hat{\pX}^{(t-1)}}}
\def\Xooh{{\hat{\pX}^{(0)}}}

\def\Xlo{{\pX^{(l-1)}}}
\def\Xli{{\pX^{(l)}}}

\def\Ti{{\pT^{(t)}}}
\def\invRo{{(\pR^{(t-1)})^{-1}}}
\def\invRoo{{(\pR^{(0)})^{-1}}}
\def\invRi{{(\pR^{(t)})^{-1}}}

\def\invRlo{{(\pR^{(l-1)})^{-1}}}

\renewcommand{\algorithmicrequire}{\textbf{Input:}}
\renewcommand{\algorithmicensure}{\textbf{Output:}}

\theoremstyle{plain}
\newtheorem{theorem}{Theorem}[section]

\theoremstyle{definition}
\newtheorem{definition}[theorem]{Definition}

\theoremstyle{remark}
\newtheorem{remark}[theorem]{Remark}

\usepackage[textsize=tiny]{todonotes}

\icmltitlerunning{Floating Anchor Diffusion Model for Multi-motif Scaffolding}

\begin{document}
\twocolumn[
\icmltitle{
Floating Anchor Diffusion Model for Multi-motif Scaffolding}



\icmlsetsymbol{equal}{*}
\icmlsetsymbol{pl}{\dag}

\begin{icmlauthorlist}
\icmlauthor{Ke Liu}{zju,equal}
\icmlauthor{Weian Mao}{adelade,zju,equal}
\icmlauthor{Shuaike Shen}{zju,equal}
\icmlauthor{Xiaoran Jiao}{zju}
\icmlauthor{Zheng Sun}{sws}
\icmlauthor{Hao Chen}{zju}
\icmlauthor{Chunhua Shen}{zju,ant}
\end{icmlauthorlist}

\icmlaffiliation{zju}{Zhejiang University, China}
\icmlaffiliation{adelade}{The University of Adelaide, Australia}

\icmlaffiliation{sws}{Swansea University, UK}
\icmlaffiliation{ant}{Ant Group}
\icmlcorrespondingauthor{Hao Chen}{haochen.cad@zju.edu.cn}

\icmlkeywords{Machine Learning, ICML}

\vskip 0.3in
]



\printAffiliationsAndNotice{\icmlEqualContribution } 

\begin{abstract}

Motif scaffolding seeks to design scaffold structures for constructing proteins with functions derived from the desired motif, which is crucial for the design of vaccines and enzymes. Previous works approach the problem by inpainting or conditional generation. Both of them can only scaffold motifs with fixed positions, and the conditional generation cannot guarantee the presence of motifs. However, prior knowledge of the relative motif positions in a protein is not readily available, and constructing a protein with multiple functions in one protein is more general and significant because of the synergies between functions. We propose a Floating Anchor Diffusion (FADiff) model. FADiff allows motifs to float rigidly and independently in the process of diffusion, which guarantees the presence of motifs and automates the motif position design. Our experiments demonstrate the efficacy of FADiff with high success rates and designable novel scaffolds. To the best of our knowledge, FADiff is the first work to tackle the challenge of scaffolding multiple motifs without relying on the expertise of relative motif positions in the protein. Code is available at \href{https://github.com/aim-uofa/FADiff}{https://github.com/aim-uofa/FADiff}

\end{abstract}

\section{Introduction}

The design of proteins with specific functions is significant for vaccines and enzymes \citep{correia2014proof,linsky2020novo,sesterhenn2020novo}. One crucial way is to design stable \textit{scaffolds} to support desired \textit{motifs} \citep{rfdiff,smcdiff,chroma}. Here motifs refer to protein structure fragments, which impart biological functions to proteins \citep{structuremotif}. Motif scaffolding has already proven to be significant in the wet experiment since drugs have been designed by solving specific instances of the motif-scaffolding problem \citep{wet1,wet2}. The development of generative models, especially diffusion models, speeds up solving the motif scaffolding problem \citep{diffusion, se3, huang2022riemannian}. Scaffolding multiple motifs in one protein is more general and significant because of the synergies between functions. However, previous works focus on scaffolding one motif at a fixed location. Adapting these approaches to scaffolding multiple motifs requires prior knowledge of relative positions between multiple motifs which is not readily available.

Previous works approach the motif-scaffold problem via conditional generation or inpainting. Conditional generation methods like SMCdiff \citep{smcdiff} are only able to scaffold one motif while the presence of motifs is not guaranteed. For inpainting methods, like Chroma \citep{chroma} and RFdiffusion \citep{rfdiff}, they fix both the structure position and sequence position of desired motifs. Consequently, to scaffold multiple motifs, the relationship between their positions must be supplied to the model in advance, necessitating domain knowledge that isn't always readily available. Even for a single motif, the sequence position is fixed.

To tackle the challenge of supporting multiple motifs, we propose a novel model dubbed \textbf{F}loating \textbf{A}nchor \textbf{Diff}usion model (\textbf{FADiff}), which not only ensures the existence of motifs but also automates motif position design without the need for prior domain expertise. The underlying principle of FADiff rests on treating the anchor motifs as rigid entities, thus permitting motifs to maintain their structures while floating. Given that a motif is composed of amino acids, it is likewise guided within the network alongside other amino acids. With the intent to preserve the structure of motifs, we treat them as rigid anchors during the diffusion process. The movement of motifs is dictated by their constituent amino acids, which further shapes the formation of the diffusion process with rigid movable substructures in this work.

Utilizing FADiff, we assure the presence of desired motifs, considering them not just as generation conditions, but as fundamental components of the generation results, analogous to the inpainting. Contrarily, while inpainting methods fixate the positions of the motifs, FADiff brings innovation to the table by independently determining the positions of each motif. 
Specifically, anchor motifs within FADiff maintain independent and rigid mobility. Guided by their internal amino acids, this property further enables the flexible movement of these anchor motifs towards rational positions.
FAdiff encourages flexible motif scaffolding, negating the need for not readily available domain expertise to assign the structural or sequential arrangement within the generated protein structure.

To demonstrate the efficacy and generalization of our FADiff, we carried out a comprehensive series of experiments. The empirical findings indicate that given multiple motifs, FADiff can effectively position them while concurrently generating designable scaffolds to support them. It is worth noting that, once trained on the task of scaffolding two motifs, FADiff can be extrapolated to scaffold any other number of motifs. These observations indicate that FADiff potentially offers a general solution to the multi-motif scaffolding problem.

To the best of our knowledge, FADiff is the first work to tackle the problem of scaffolding multiple motifs without the need for prior knowledge of the relative positions of multiple motifs, which is often unobtainable. 
The main contributions of our work can be summarized as follows:

\begin{itemize}
    \item We propose a practical and significant problem of scaffolding multiple motifs where the prior knowledge of their relative positions is not readily available.
    \item We propose a new diffusion model, floating anchor diffusion (FADiff) to tackle the problem of scaffolding multiple motifs. FADiff assures the existence of motifs and automates the design of motif position by facilitating the rigid movement of the motifs.
    \item Our experiments demonstrate that FADiff can float the anchor motifs to rational positions and generate designable scaffolds to support them. The generalization of FADiff indicates its potential to be a general solution to multiple motif scaffolding.
\end{itemize}

\section{Related Works}
\subsection{Motif scaffolding problem}
Multi-motif scaffolding is a central task in protein design. For example, a protein boasting high specificity can be fashioned by assimilating several recognized binding motifs \citep{motifscaffold,cao2022design,jiang2023novo}. Furthermore, via expert knowledge, a pair of EF-hand motifs are effectively merged into the protein structure \citep{efhand}. Importantly, in many instances, either sequence or structure relative positions between motifs remain undetermined. While this situation permits the resolution of some issues, it persistently demands considerable experimentation, human intervention, and specialized knowledge \citep{roel2023single,davila2023directing,roy2023novo}. Additionally, these strategies display pronounced shortcomings, particularly when confronting conditions devoid of suitable templates and references in the Protein Data Bank (PDB) \citep{pdb}. FADiff provides a general solution to multi-motif scaffolding without any reliance on domain expertise.

\subsection{Generative models for scaffolding motifs}
The advent of generative protein models \citep{riemannian, lee2023score, progen, smcdiff, gruver2023protein,lisanza2023joint} has instigated a dramatic evolution in protein design. Motif-scaffolding, a pivotal undertaking within protein design, has been consistently broached by diverse diffusion model techniques throughout the years. Generative models try to solve the motif scaffolding problem by conditional generation or inpainting. For example, SMCDiff \citep{smcdiff} and Chroma\citep{chroma} take motifs as guidance for their pre-trained unconditional model to generate proteins with motifs in it. RFdiffusion \citep{rfdiff} fixes the motifs in a protein and paints the scaffold. 
However, both two methods fail to scaffold multiple motifs since the motif positions in the protein are manually determined and fixed for them. The conditional generative methods even cannot guarantee the presence of motifs in the generated protein.
FADiff solves the problem by enabling the motifs to float rigidly in the diffusion process, which leads to the automatic position design and the existence of motifs in the generated protein.

\section{Preliminaries and Notation}
\subsection{The multi-motif scaffolding problem}
A protein $\pP = \{\pA, \pX\}$ is defined by its amino acid sequence $\mathcal{A}=\{a_1,a_2,\cdots, a_n\}$ and backbone structure $\pX = [\bx_1, \bx_2, \cdots, \bx_n]^T \in \mathbb{R}^{n,3}$, where $n$ denotes the number of amino acids in a protein. $a_i \in \mathcal{C}^{20}$ denotes the type of $i$-th amino acids, where $\mathcal{C}$ is a set of 20 genetically-encoded amino acids. $\bx_i \in \mathbb{R}^3$ is the $i$-th C-$\alpha$ residue backbone coordinates in 3D. The 3D structure of a protein can be determined by its corresponding amino acid sequence, \ie, $\pX(\mathcal{A})$. In addition, the order of the amino acids in the sequence, \ie, the sequence position, is also an important piece of implicit information in $\mathcal{A}$, denoted by $\mathcal{D}$. Thus, the amino acid sequence consists of the sequence position and amino acid types, \ie, $\mathcal{A} = \{\mathcal{C}^{n}, \mathcal{D}\}$. We can define a protein as:
\begin{definition}[Protein structure]
    A protein structure consists of amino acid sequence $\mathcal{A}$ and backbone structure $\pX$, where $\mathcal{A}$ contains both the amino acid types $\mathcal{C}$ and index in sequence $\mathcal{D}$, \ie, $\pP = \{\mathcal{A}, \pX\} = \{\mathcal{C}^{n}, \mathcal{D}, \pX\}$
\end{definition}
Given a protein $\pP$, we can divide it into the functional motif $\pM$ and the scaffold $\pS$. Since multiple motifs exist in one protein, $\pM = \{ \pM_1, \pM_2, \cdots, \pM_m \}$, the protein is denoted as $\pP = \pM_\pP \cup \pS_\pP = \{\pM_1, \pM_2, \cdots, \pM_m, \pS\}$, where $m$ is the number of motifs. Therefore, we can define the scaffolded motif and scaffolding as:
\begin{definition}[Scaffolded motif and scaffold]
    We can describe a scaffolded motif $\pM_\pP$ in a protein with its positions in the protein and its internal structure, specifically, the internal structure $\pM_{\rm X}$, the internal sequence $\pM_\pA$, the position in the protein structure $\pX_\pM$, and the position in the protein sequence $\pA_\pM$ of motif, \ie, $\pM_\pP = \{\pM_{\rm X}, \pM_\pA, \pX_\pM, \pA_\pM\}$. Similarly, the scaffolding can be defined as $\pS_\pP = \{\pS_\pX, \pS_\pA, \pX_\pS, \pA_\pS\}$
\end{definition}

\begin{figure*}[t]
    \centering
    \includegraphics[width=0.98581\linewidth]{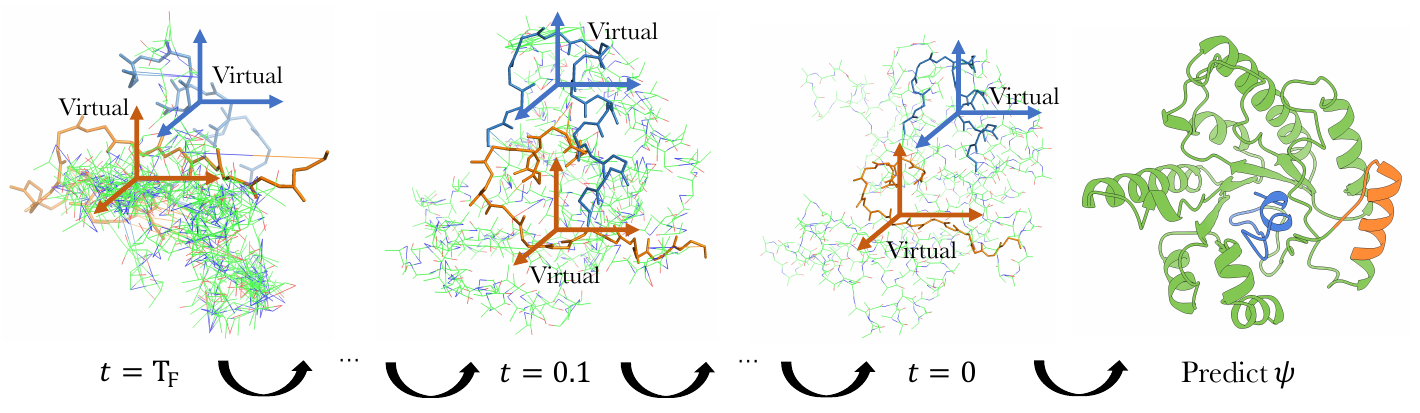}
    \caption{
    In the denoising process, we keep the motifs translating and rotating rigidly, which means the internal structure of motifs is maintained while their positions in the protein are flexible. The \textcolor{orange}{orange} and \textcolor{blue}{blue} colors indicate the anchor motifs that float rigidly. The \textcolor{green}{green} color indicates the scaffold residues. The coordinate system in color denotes the virtual coordinate system, which is the geometry center of each motif.}\label{fig:frame}
\end{figure*}

The common motif-scaffolding settings focus on the backbone generation and the order of residues $\pD$ in protein amino acids sequence $\pA$ is considered. Therefore we ignore the $\mathcal{C}$ in $\pA$ in the below as per the common setup.
Inpainting methods \cite{chroma, rfdiff} for motif scaffolding require the motif structure, its position in the protein structure, and its sequence position in the protein sequence prior, which can be remarked as:
\begin{remark}[Motif-scaffolding by inpainting]
    \label{remark:inpainting}
    Inpainting methods seek to predict the scaffolding structures $\pS_\pP$ given the motif structure $\pM_\pP$, \ie, $\pS_\pP = f(\pM_{\rm X}, \pM_\pA, \pX_\pM, \pA_\pM)$.
\end{remark}
The motif position in the protein $\pX_\pM, \pA_\pM$ is specified manually in inpainting methods, which is not readily available.

The conditional generation methods \cite{smcdiff} require only the motif internal structures to predict the structure of the whole protein, which can be remarked as:
\begin{remark}[Motif-scaffolding by conditional generation]
    Conditional generation methods sought to predict the protein structures $\pP$ given the motif's internal structure, \ie, $\pP = f(\pM_{\rm X}, \pM_\pA)$.
\end{remark}
In the conditional generation, the presence of motifs is not guaranteed. For multiple motifs which are encoded together, the relative position between them is fixed.

To maintain the motif in the generated protein and enable the automatic position design, we formulate the multiple motif scaffolding problem as:
\begin{definition}[Multiple motif scaffolding problem]
    Given the internal structure of multiple motifs, the multiple motif scaffolding seeks to predict the scaffolding and the motif positions in the protein, \ie, $\{\pS_\pP, \pX_\pM, \pA_\pM \} = f( \pM_{\rm X}, \pM_\pA )$
\end{definition}

\subsection{Backbone parameterization}
We adopt the protein backbone parameterization and notations in FrameDiff \citep{se3}. Each residue backbone is parameterized by an orientation preserving rigid transformation (\textit{frame}) $\pT \in \mathbb{R}^{4\times4}$ that maps from fixed coordinates ${\rm N^*, C_\alpha^*, C^*, O^*} \in \mathbb{R}^3$ centers at ${\rm C_\alpha^*}=(0,0,0)$.
Thus, the main atom coordinates of the $i$-th residue on the backbone are obtained as
\begin{equation*}
    {\rm [N_\textit{i}, C_\textit{i}, (C_\alpha)_\textit{i}]} = \pT_i \cdot {\rm [N^*, C_\alpha^*, C^*]},
\end{equation*}
where $\pT_i$ is an operation of the special Euclidean (SE(3)) group. Each transformation $\pT_i$ can be decomposed into rotation $\pR_i \in \mathbb{R}^{3\times 3}$ and translation $\pX_i \in \mathbb{R}^3$, \ie, $\pT_i = (\pR_i, \pX_i)$, where $\pR_i \in {\rm SO(3)}$. Therefore, given a coordinate ${\bf v} \in \mathbb{R}^3$ in the $i$-th frame, its location in the fixed coordinate is given as 
\begin{equation}
    \pT_i {\bf v} = \pR_i{\bf v} + \pX_i.
    \label{equ:transform}
\end{equation}
Further, with an additional torsion angle $\phi$, the coordinates of atom O in the residue can be determined.
Different from FrameDiff, FADiff takes each motif as rigid and enables their movement in the diffusion process, as shown in Fig.~\ref{fig:frame}, \ie, the structure of each motif is preserved and can float rigidly. 
The movement of each motif is steered by the average of its constituent residues. To get the translation of each residue caused by the rigid anchor motif rotation, we define a virtual coordinate system with a rotation matrix of identity $I$ and a translation of $\pX_v = \overline{\pxm}$.

\subsection{Diffusion model for protein backbone generation}
We follow the Riemannian score-based generative modeling of \citet{riemannian} and \citet{se3}. Denoising score matching (DSM) aims to approximate the Stein score $\nabla \log p_t(x)$, which is unavailable in practice, with a score network $s_\theta(t, \cdot)$ through minimizing the DSM loss:
\begin{equation*}
    \label{eq:scorematch}
    \mathcal{L}(\theta) = \mathbb{E}\left[\lambda_t \Vert \nabla \log p_{t\vert 0}(\dX^{(t)} \vert {\dX}^{(0)}) - s_\theta(t,\dX^{(t)}) \Vert^2 \right],
\end{equation*}
where $\dX$, $p_{t\vert 0}$, $\lambda_t > 0$, and $\theta$ denote the data distribution, the density of $\dX^{(t)}$ given $\dX^{(0)}$, a weight, and the network parameters, respectively. The expectation $\mathbb{E}$ is over the $t \sim \mathcal{U}([0, T_F])$ and $(\dX^{(0)}, \dX^{(t)})$.

\begin{figure*}[t!]
    \centering
    \includegraphics[width=0.98581\linewidth]{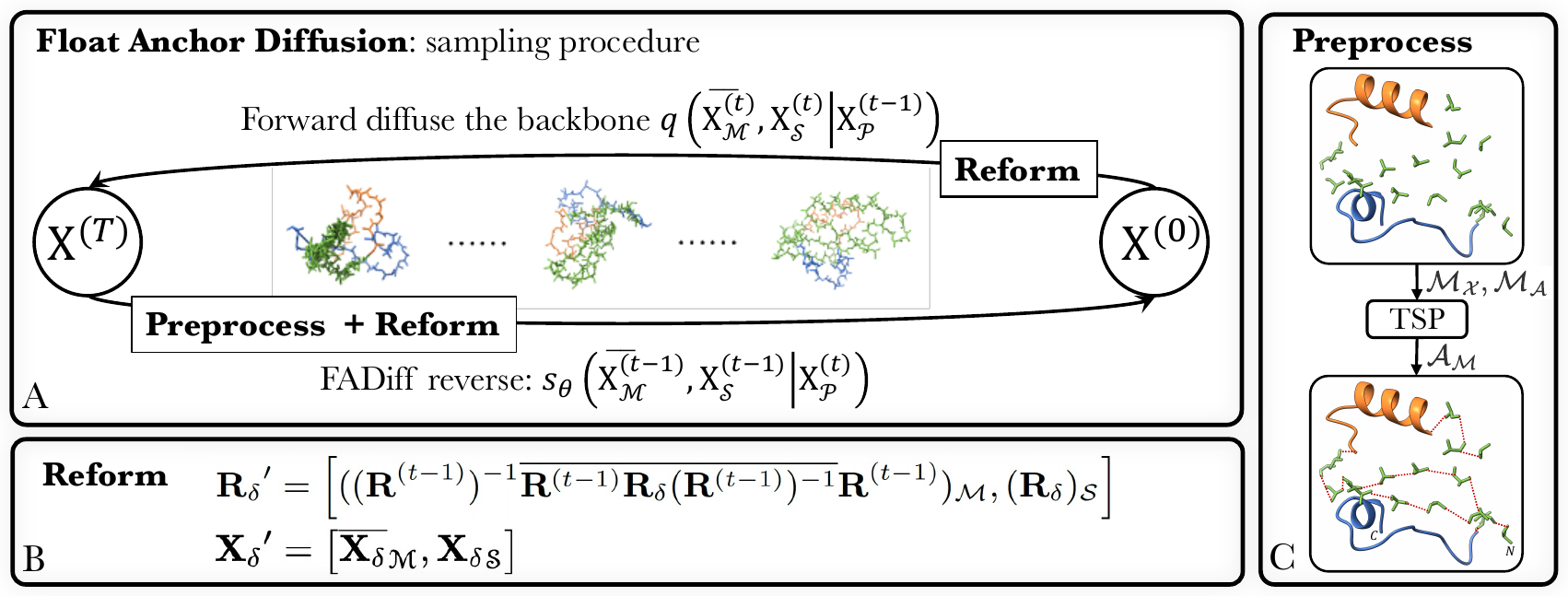}
    \caption{\textbf{A)} Given multiple motifs with their internal structure $\pM_{\rm X}$ and $\pM_\pA$. We specify the sequence position of residues by finding the shortest chain with a greedy algorithm like the traveling salesman problem (TSP), where the distances are the gaps between the atoms C and N of two residues. In both the forward process and reverse process, we take the motifs as rigid and enable them to float rigidly. \textbf{B)} Generally, we reform the noise and updates for each motif. \textbf{C)} The preprocess of TSP. The \textcolor{orange}{orange} and \textcolor{blue}{blue} colors indicate the motifs. The residues in \textcolor{green}{green} are generated scaffolds.}
    \label{fig:model}
\end{figure*}

\subsection{Additional notations}
The motif and scaffold parts of proteins are denoted by subscript $\pM$ and $\pS$ respectively. $\pR=\{\pR_\pM, \pR_\pS \}=\{{\pR_\pM}_1, {\pR_\pM}_2, \cdots, {\pR_\pM}_m, \pR_\pS\}$ and $\pX=\{\pX_\pM, \pX_\pS \}=\{{\pX_\pM}_1, {\pX_\pM}_2, \cdots, {\pX_\pM}_m, \pX_\pS\}$ denote the rotation and translation of all the residues in a protein. $\pT=\{\pR, \pX \}$ denotes the position of residues. We denote noise, perturbation, and update by the notation with a $\delta$ subscript. For example, the noise to the rotation and translation is denoted as $\rn$ and $\tn$ respectively. The average operation $\overline{\cdot}$ over motifs indicates the average over each motif part, respectively. $[\cdot, \cdot]$ indicates the two elements, \ie motif elements and scaffold elements.

\section{Floating Anchor Diffusion}
To tackle the challenge of generating scaffolds to support multiple motifs without prior knowledge of their relative positions, we propose a Floating Anchor Diffusion model (FADiff) as shown in Fig.~\ref{fig:model}. 
The core concept of FADiff lies in treating the motifs as rigid and enabling them to move rigidly. The motivation can be summarized as (1) For multiple motifs, their relative positions cannot be determined manually. Therefore, FADiff allows them to float independently in the diffusion process. (2) Since motif scaffolding needs the presence of motifs in the designed protein, FADiff preserves the structure of each motif as anchors rigidly. (3) Motifs are composed of amino acids, thus the movements are determined by their internal amino acids. Generally, we average the movement of residues inside each motif to steer it and make the whole process consistent with the diffusion process as shown in Fig.~\ref{fig:frame}.

\subsection{Forward diffuse the protein backbone}
To model the forward diffusion process on protein backbone, $q(\dX_\pM^{(t)}, \dX_\pS^{(t)} \vert \dX_\pP^{(t-1)})$, we add noise to the frames following FrameDiff \citep{se3} but average the noise on motifs for treating the motif as rigid. We divide the transformation of frames into rotation and translation.
\subsubsection{Rotation}
For a randomly sampled SO(3) rotation noise $\rn$ to rotation $\Ro$, we estimate the movement of motifs with the average of their constituent residues, \ie, the noise is reformed as:
\begin{align}
    \label{eq:noise1r}
    \rn' = \big[(\invRo\overline{\Ro\rn\invRo} & \Ro)_\pM, \notag\\
    &(\rn)_\pS \big]
\end{align}
The two items indicate the motif part and scaffold part respectively. The average operation $\overline{\cdot}$ over motifs indicates the average over each motif part respectively. The reformed noise in Eq.~\eqref{eq:noise1r} is obtained as follows. With a randomly sampled noise $\rn$, in order to maintain the rigidity of motifs, we estimate the movement of motifs with the average value of their constituent residues. Specifically, we define a virtual frame at the geometry center of motifs with a rotation matrix of identity $I$ and a translation of $\pX_v = \overline{\pxm}$. Then we apply the noise to each residue to get the possible transformation of frames $\Ri'=\Ro\rn$. The rotation transformation between the original and transformed motif in the virtual coordinate system is $\Ri'\invRo = \Ro\rn\invRo$. To estimate the rotation of anchor motifs efficiently, we average the quaternion of its constituent residues, \ie,
\begin{align}
    \label{eq:delta}
    [&\Delta\pR_\pM, \Delta\pR_\pS] =  \\
    &\left[(\overline{\Ro\rn\invRo})_\pM, (\Ro\rn\invRo)_\pS\right], \notag
\end{align}
Without loss of generalization, any other rotation average methods can be applied here to estimate the rotation of motifs. Finally, we transform the rotation of the anchor motif under the coordinates of the virtual frame back to each residue as

\begin{align}
\label{eq:rotationri}
    \Ri = \left[\Delta\pR_\pM, \Delta\pR_\pS\right] \Ro.
\end{align}
The transform from $\Ro$ to $\Ri$ \ie, the noise actually added to $\Ro$ is:
\begin{equation}
    \label{eq:actualnoise}
    \rn' = \invRo[\Delta\prm, \Delta\prs] \Ro,
\end{equation}
which is used to calculate the rotation score. The details can be found in Appendix \ref{app:noise}.
To get the translation of motifs' constituent residues, we first estimate the translation of their constituent residues caused by their rotations, which can be obtained as:
\begin{equation}
    \label{eq:forward}
    \Delta \pX_\pM = \Delta\prm (\Xo_\pM - \pX_v) + \pX_v - \Xo_\pM
\end{equation}

\subsubsection{Translation}
Given a randomly sampled noise on translation $\tn$, we also average the noise on each motif, respectively, as 
\begin{equation}
    \tn' = \left[\overline{\tn}_\pM, \tn_\pS \right],
\end{equation}
which is used for translation score calculation. Then $\Xi$ is obtained through
\begin{equation}
    \Xi = \Xo + \tn'.
\end{equation}

Finally, the noised data $\pT$ for the score network to denoise is 
\begin{equation}
    \Ti = (\Ri, \Xi+[\Delta \pX_\pM, 0_\pS])
\end{equation}

\subsection{Denoising score matching}
Given $\Ti$, the score network is designed to conduct iterative updates on the frames across a sequence of $L$ layers, eventually yielding the predicted protein position $\hat{\pT}^{(t-1)}$ \citep{af2,vfn}. Then the score is calculated with $\Ti$ and $\hat{\pT}^{(t-1)}$. To keep the motifs rigid, we average the update for motifs like that in the forward diffusion process. When calculating the score, we remove the residue translations caused by the rigid rotation to keep consistent with the diffusion process.
\subsubsection{Frame update}
Similar to the process in forward diffusion, we reform the predicted update $\rnh$ on $\Rlo$ as:
\begin{align}
    \label{eq:frameupdate}
    \rnh'= \left[(\invRlo\overline{\Rlo\rnh\invRlo}\right. & \Rlo)_\pM,   \notag \\ 
    &\left. (\rnh)_\pS \right]
\end{align}

we first estimate the rotation of each anchor motif with the average of its internal residue rotation. Then the rotation of each residue can be obtained as 
\begin{align}
    \label{rq:lastlayer}
    &\Rli = \left[\Delta\hat{\pR}_\pM, \Delta\hat{\pR}_\pS \right] \Rlo,\\
    [&\Delta\hat{\pR}_\pM, \Delta\hat{\pR}_\pS ]= \notag \\
    &\left[ (\overline{\Rlo\rnh\invRlo})_\pM, (\Rlo\rnh\invRlo)_\pS \right].\notag
\end{align}
The translation of each residue caused by the rotation is:
\begin{align}
    \label{eq:rotationcause}
    \Delta \hat{\pX}_\pM = \Delta\hat{\pR}_\pM (\Xlo_\pM - \pX_v) + \pX_v - \Xlo_\pM.
\end{align}
Then the $\Xli$ is derived with Eq.~\eqref{equ:transform} as:
\begin{align}
    \label{eq:xl}
    \Xli =& \left[\overline{(\Rlo\tnh)_\pM-\Delta\hat{\pX}_\pM}+\Delta\hat{\pX}_\pM, (\Rlo\tnh)_\pS\right] \notag \\
    &+ \Xlo
\end{align}
The details of Eq.~\eqref{eq:xl} can be found in the Appendix \ref{app:Translation}.
We adopt VFN-Diff \citep{vfn}, a SE(3) diffusion protein structure generation model as our score network in this work. Without loss of generalization, any other SE(3) diffusion model can be used as a score network here.

\subsubsection{Score calculation}
To keep consistent with the score-based diffusion process \citep{diffusion, se3}, we remove the residue translations caused by the rigid anchor motif rotation for the score calculation. Similar to the process above, we have the translation caused by rigid anchor motif rotation as:
\begin{equation}
\label{eq:transupdaterota}
\Delta \pX_\pM' = (\Ri)^{-1} \hat{\pR}^{(0)} (\Xi - \pX_v) + \pX_v - \Xi,
\end{equation}
where $\hat{\pR}^{(0)} = \Rli$ is the prediction of the network.
Finally, the rotation and translation for score calculation are $\Rli$ and $\Xli-[\Delta \pX_\pM', 0_\pS]$. The details can be found at Appendix~\ref{app:score_trans}.

\subsection{Training loss}
With the DSM loss in Eq.~\eqref{eq:scorematchloss}, the scheduler for rotation as $\lambda_t^r = 1/\mathbb{E}[\Vert \log p_{(t\vert 0)}(\pR_n^{(t)}\vert \pR^{(0)}) \Vert_{SO(3)}^2 ]$, and the scheduler for translation as $\lambda_t^{x}=(1-e^{-t}/e^{-t/2})$ following \citet{se3}, we have the DSM loss as:
\begin{equation*}
    \label{eq:scorematchloss}
    \mathcal{L}_{dsm} = \mathbb{E}\left[\lambda_t \Vert \nabla \log p_{t\vert 0}(\dX^{(t)} \vert {\dX}^{(0)}) - s_\theta(t,\dX^{(t)}) \Vert^2 \right],
\end{equation*}
which is consistent with the score-based diffusion model. More details can be found in the Appendix \ref{app:consistent}.

Since the broken C-N bonds are found in early experiments, we have two auxiliary losses to get the distance between the atom C and N of two residues into the right range with Eq.~\eqref{eq:c_n_loss} and to get the atoms in the right place with Eq.~\eqref{eq:bb_loss} following \citep{af2,se3} as follows:
\begin{align}
    \label{eq:c_n_loss}
    &\mathcal{L}_{c-n} = \frac{1}{4n} \sum_{i=1}^{n} \sum_{\bx\in\Omega} \|\bx_i^{(0)} - \hat{\bx}_i^{(0)}\|^2,\\
    \label{eq:bb_loss}
    \mathcal{L}_{bb} &= \frac{1}{Z} \sum_{i,j=1}^{n} \sum_{a,b\in\Omega} \mathbbm{1}\{d_{ab}^{ij} < 0.6\} \| d_{ab}^{ij} - \hat{d}_{ab}^{ij} \|^2,  \\
    Z &= \left( \sum_{i,j=1}^{n} \sum_{a,b\in\Omega} \mathbbm{1}\{d_{ab}^{ij} < 0.6\} \right) - n ,\notag
\end{align}
where $\Omega$ is the set of atoms $\{\rm C, C_\alpha, O, N \}$. $d_{ab}^{ij}$ and$\hat{d}_{ab}^{ij}$ indicate the ground truth and predicted distance between atom $a$ and $b$ in residue $i$ and $j$. With $\mathbbm{1}\{d_{ab}^{ij} < 0.6\}$, we leave alone the distances larger than 0.6\text{\AA}. For more details in Appendix \ref{app:loss}

\subsection{Sampling}
Euler-Maruyama discretization with 500 steps implemented as a geodesic random walk is adopted in this work following \citet{riemannian,se3,rfdiff}. 
In this work, the sequence is constructed by finding the shortest chain, like the traveling salesman problem (TSP) where the distance is the gap between atoms C and N of two residues.
More details are in the Appendix~\ref{app:sample}.

\section{Experiments}
We trained FADiff on the task of scaffolding two motifs of lengths from 20 to 80 residues with the virtual motif (VM) dataset. We first analyze the performance of FADiff on the evaluation set of the VM dataset. Then we evaluate the performance and generalizability of FADiff on the multi-motif scaffolding (MS) Benchmark and analyze the generated samples in terms of designability. An ablation study is conducted to evaluate the effectiveness of TSP and noise scale. Finally, we compare our approaches with the conditional generation and inpainting methods which further demonstrate the efficacy of FADiff.

\subsection{Setup}
\paragraph{Dataset.}
Two datasets are utilized in this work, including the virtual motif dataset (\textbf{VM dataset}) from the PDB database \citep{pdb} for training and the evaluation multi-motif scaffolding benchmark \textbf{MS Benchmark} that we collected from the PROSITE database\citep{PROSITE}. \textbf{VM dataset} contains 59,128 proteins with chain lengths from 60 to 512 residues extracted from the PDB database. For each entry, we randomly crop two fragments with lengths of 20 to 80 residues from the protein as virtual motifs for training. \textbf{MS Benchmark} contains 16,251 functional motifs with lengths between 10 and 20 residues \citep{xiong2006essential} that naturally exist.

\paragraph{Evaluation metrics.}
We mainly employ the self-consistence TM-score (\textbf{scTM}) to evaluate the \textit{designablitity} of generated structures and the \textbf{\textit{in silico} Success Rate} (\textbf{SR}) to

evaluates the performance of the model following previous works \citep{smcdiff,tmscore,chroma}. A higher TM-score or scTM indicates two structures are more similar. 
\textbf{scTM} is the TM-score between the generated structures and the reconstructed structure through ProteinMPNN \citep{proteinmpnn} and ESMFold \citep{esmfold} as shown in Fig~\ref{fig:evaluation}. The generated structure is \textit{designable} if $\rm scTM > 0.5$. 

\textbf{SR} is the ratio of designable structures in the generation. Following previous work \citep{smcdiff,tmscore}, for each group motif to be scaffolded, we generate 5 samples for each length of 160 to 410 residues and run ESMFold 8 times to get the highest scTM.

\textbf{Motif RMSD}, \ie, the difference between the desired motif and corresponding structure in the generated protein, used in conditional generation methods \citep{smcdiff} to evaluate the presence of motifs is not applicable here since it is \textit{0} for our model consistently. Details can be found in  Appendix \ref{app:evaluation}.

\paragraph{Compared approaches.}
We mainly compare FADiff with \textit{conditional generation} methods \citep{smcdiff} and \textit{inpainting} methods \cite{rfdiff,chroma}. We adapt inpainting methods to multiple motif scaffolding by randomly assigning the relative structure position of two motifs since they require positions of motifs as input as \textit{Remark} ~\ref{remark:inpainting}. Due to the probable absence of motifs in the conditional generation, one generation is successful if (1) $\rm scTM>0.5$ and (2) the $\rm motif RMSD < 0.1$ following previous works \citep{smcdiff}.
\begin{figure*}[t!]
    \centering
    \includegraphics[width=1\linewidth]{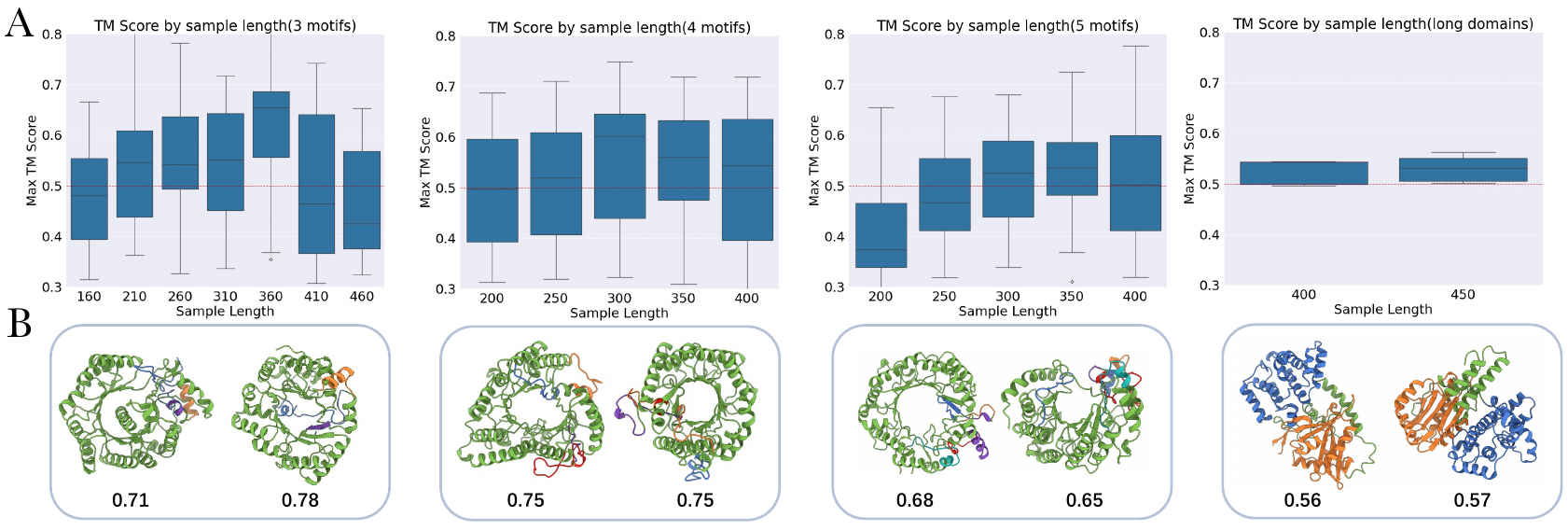}
    \caption{Statistic analysis and visualization of generation results for scaffolding 3, 4, 5, and two huge domains of length more than 100 residues. \textbf{A)} scTM distribution. The samples over the red dashed line are designable. 59.18\%, 46.00\%, 36.15\%, and 60.00\% generated protein structures are designable for scaffolding 3, 4, 5, and two huge domains. \textbf{B)} Generated protein structures. The \textcolor{green}{green} colors indicate the generated scaffolds and the other colors indicate the motifs. The numbers below each generated structure indicate the scTM score.}
    \label{fig:motif345huge}
\end{figure*}
\begin{figure}[t!]
    \centering
    \includegraphics[width=0.98581\linewidth]{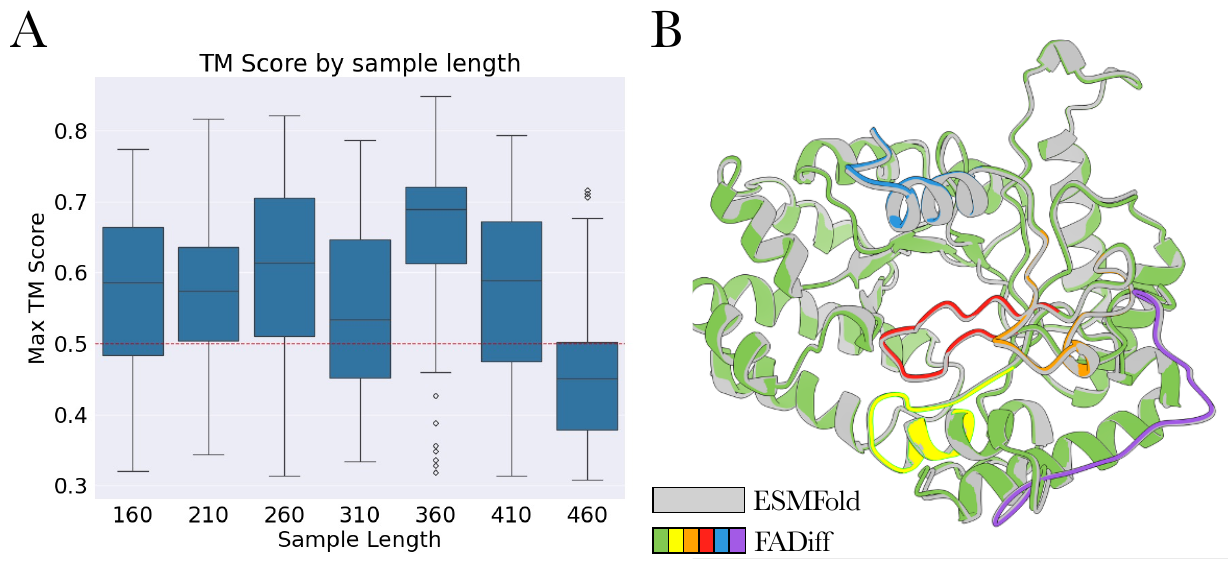}
    \caption{\textbf{A)} The distribution of scTM for inference results with varying lengths. The samples above the red dashed line are designable. 
    \textbf{B)} visualization results were obtained by training the model on two motifs and testing on five motifs. `ESMFold' denotes protein structures constructed through the ProteinMPNN and ESMFold, with a preference for closer structural resemblance. Our model demonstrates generalization capability.}
    \label{fig:2motif}

\end{figure}

\subsection{Experimental results}
The evaluation of FADiff on the VM dataset demonstrates its ability to generate designable protein structures with a high tm-score. Experiments on the MS Dataset for two motifs demonstrate the efficacy of FADiff with a high ratio of designable structures in the generated samples. To evaluate the generalizability of FADiff, experiments on scaffolding more than two motifs are conducted with FADiff trained on the two-motif scaffolding task. Finally, an ablation study is conducted to demonstrate the effectiveness of the choice in implementation.
\subsubsection{Evaluation on VM dataset}

The TM-score between the protein structure where the virtual motifs are located and the structures generated by FADiff in the evaluation set of the VM dataset is shown in Fig.~\ref{fig:tmscore}. 
The TM-score of 67.5\% generated structures is above 0.5 which indicates that the generated structure and the original structures are similar, demonstrating the ability of FADiff to generate designable protein structures. Besides, the novelty of generated proteins provides insight into a way to design novel proteins, \ie, scaffolding motifs, as shown in Table \ref{tab:pdbtm}. More details can be found in Appendix~\ref{app:tmscore_eval}.

\subsubsection{Evaluation on MS Benchmark}
\paragraph{Scaffold two motifs.}
We evaluate FADiff on the task of scaffolding two functional motifs from the MS Benchmark. The ratio of designable protein structures generated by FADiff is 73.05\% as shown in Fig.~\ref{fig:2motif}A. The performance of FADiff varies with the length of the generated protein due to the bias of training data. The distribution of proteins varies with their lengths in the VM dataset.
FADiff also demonstrates commendable performance on scaffolding 5 motifs, as depicted in Fig.~\ref{fig:2motif}B.

\paragraph{Generalization of FADiff.} To evaluate the generalization of FADiff, we apply FADiff trained on the task of scaffolding two virtual motifs to scaffold 3, 4, 5, and two huge domains as shown in Fig.~\ref{fig:motif345huge}.
(1) \textbf{Scaffold more than two motifs}:
The average \textit{in silico} success rates of FADiff for scaffolding 3, 4, and 5 motifs achieve 62.38\%, 58.40\%, 46.67\%.
(2) \textbf{Scaffold huge domains}:
We further design scaffolds for two huge domains with lengths of more than 100 residues which are also never trained with. Domains are also functional parts of proteins. The average \textit{in silico} success rate for two huge domains achieves 80.00\%, which further demonstrates the generalization of FADiff. The generalization is achieved since FADiff treats all the residues equally. The decrease in success rate for scaffolding more motifs is caused by the increasing difficulty, especially with fewer scaffold residues.

\subsection{Ablation study}

\begin{table}[t!]
    \centering
    \caption{\textit{In silico} success rate (\%) for different lengths, with/without TSP, and different translation noise scales in sampling. TSP and Random indicate the preprocessing method. $/$2 and $\times$2 indicate the translation noise scale in sampling.}
    \label{tab:ablation}
    \begin{tabular}{clccccc}
        \toprule[1.2pt]
        Method& 160 & 210 & 260 & 310 & 360 \\
        \midrule 
        FADiff  &\textbf{76.67} & 71.67 & \textbf{81.67} & 65.00 & \textbf{86.67}\\
        Random  & 49.33 & 60.00 & 60.00 & 54.67 & 82.67\\ 
        $/$2  &  69.23 & \textbf{76.92} & 76.92 & 60.00 & 83.08 \\
        $\times$2 & 70.00 & 70.00 & 75.00 & \textbf{78.33} & 83.33 \\ 
        \bottomrule[1.2pt]
    \end{tabular}
\end{table}
\paragraph{Noise scale on translation.}
Since the motifs are much bigger than residues, it is straightforward to increase the translation noise to enable the residues of scaffolds to navigate in the scale of motifs. We train a model without increasing the noise scale on translation and test it on scaffolding two motifs. The average scTM is 0.213 and the average SR is close to $0\%$ due to significant translation prediction errors. Please refer to Appendix \ref{app:hyper-param} for more details.

\paragraph{TSP for sequence construction.}
To evaluate the efficacy of TSP, we randomly connect the residues to construct the amino acid sequence. Although TSP outperforms the random connection consistently, the random connection also leads to a high average \textit{in silico} success rate of 60.67\% as shown in Table \ref{tab:ablation}. More details are in Appendix \ref{app:tsp}.

\paragraph{Noise scale in sampling.}
With different noise scales on translation in sampling, the performance of FADiff varies little and achieves a high average \textit{in silico} success rate of 72.56\% and 72.22\% for the reducing by half ($/$2) and augmenting by twice the noise scale ($\times$2).

\subsection{Comparison}
We compare FADiff with conditional generation and inpainting methods. The inpainting is adopted to scaffold multiple motifs by randomly assigning their relative positions. One generation is successful for the conditional generation method if the motif RMSD is less than $1$ and $\rm scTM>0.5$ following \citet{smcdiff}. 
FADiff outperforms inpainting and conditional generation consistently in the success rate of scaffolding two motifs over all different lengths of proteins as shown in Table \ref{tab:compare}.
The generated structures by conditional generation have high scTM scores while the existence of desired motifs is not ensured. 87.25\% of the generated structures' scTM scores are over 0.5, indicating they are designable. However, the motif RMSD of only 21.50\% generated structures is under 0.1, which indicates the absence of desired motifs in the generation. 
Inpainting outperforms the conditional generation methods since the presence of motifs in the generation is guaranteed. However, the randomly assigned inappropriate relative positions of motifs lead to the failure in the generation. With FADiff, we ensure the existence and automate the design of motif relative positions by enabling the motifs to float rigidly.

For more results on specific motifs and case studies, please refer to Appendix \ref{app:Results}.
\begin{table}[t!]
    \centering
    \caption{\textit{In silico} success rate (\%) for different lengths with different scaffolding methods, where Condition indicates the conditional generation method.}
    \label{tab:compare}
    \begin{tabular}{clccccc}
        \toprule[1.2pt]
        Method& 160 & 210 & 260 & 310 & 360\\
        \midrule 
         \cellcolor{lm_purple_low}FADiff  &\cellcolor{lm_purple_low}\textbf{76.67} & \cellcolor{lm_purple_low}\textbf{71.67} & \cellcolor{lm_purple_low}\textbf{81.67} & \cellcolor{lm_purple_low}\textbf{65.00} & \cellcolor{lm_purple_low}\textbf{86.67}\\
        Inpainting  & 51.58 & 56.84 & 62.11 & 60.00 & 79.47 \\ 
        Condition  &  23.75 & 22.50 & 21.25 & 23.75 & 16.25 \\
        \bottomrule[1.2pt]
    \end{tabular}
\end{table}
\section{Discussion}
\paragraph{Why FADiff operate effectively without the expertise on the relative positions of multiple motifs?} Previous works fail to scaffold multiple motifs due to the strong correlation between sequence position and structure position. In previous works, like RFDiffusion, the positions of multiple motifs on the amino sequence $\pA$ and in the protein structure $\pX$ should be specified manually. Since the protein structure $\pX$ is determined by the amino sequence $\pA$, \ie $\pX(\pA)$ and $\pX_\pM(\pA_\pM)$, the manually assigned positions are almost unable to achieve the correlation. However, FADiff allows the anchor motifs to float to a rational position as a rigid, enabling the automatic design of relative positions of multiple motifs. Therefore, even with a random connection to construct the amino acid sequence, FADiff also achieves a high success rate by steering the motifs to a rational position determined by the sequence.

\paragraph{How does FADiff generalize to scaffold multiple motifs?}

In both the diffusion and reverse processes, all the residues of motifs and scaffolds are considered equally for FADiff. Only for the update of residue positions, we average the movement of motif residues to steer the motifs. The whole process can be considered consistent with the diffusion model for protein backbone generation. A FADiff trained on scaffolding two virtual motifs of lengths from 20 to 80 can be applied to scaffolding more than two motifs and huge domains with a length of more than 100 residues. 

\section{Conclusion}
To tackle the challenge of automatically designing relative positions and preserving the presence of multiple motifs. We propose a Floating Anchor Diffusion (FADiff) model for scaffolding multiple motifs for the first time. FADiff solves the problem by taking the anchor motifs as rigid respectively and allowing them to float flexibly. Our experiments on the benchmark demonstrate the efficacy and generalization of FADiff, providing insights for future wet experiments and a new way to construct novel protein structures. It is straightforward to apply FADiff to other generation tasks where multiple substructures should be preserved while their positions in the generation are flexible.

\section*{Impact Statement}
The goal of this work is to advance the field of Machine Learning and Computational Biology. While there are many potential societal consequences of our work, we believe that none of which must be specifically highlighted here.

\bibliography{example_paper}
\bibliographystyle{icml2024}

\newpage
\appendix
\onecolumn
\section{Notations}
\begin{figure}[ht!]
    \centering
    \includegraphics[width=0.7\linewidth]{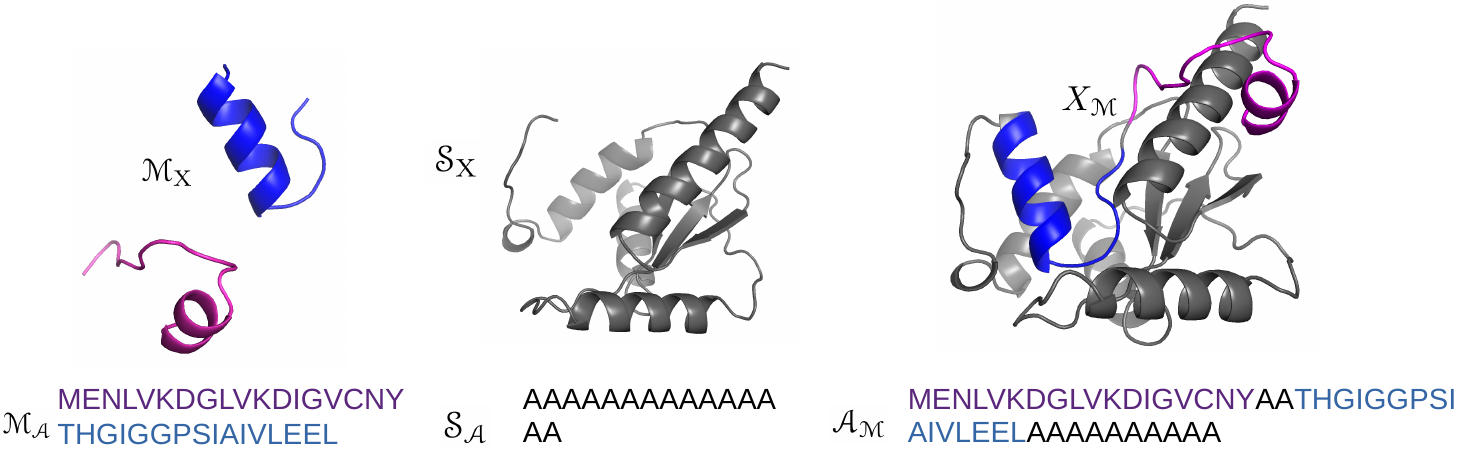}
    \caption{Illustration of notations}
    \label{fig:notation}
    \vspace{-0.4cm}
\end{figure}

The notations in this paper follow the principle that $\mathcal{M}$ with a subscript describes the structure of Motifs solely and $\mathcal{M}$ as a subscript describes the position or structure of motifs in the designed protein. The notations used in this paper are described in Table.~\ref{tab:notation} and Fig.~\ref{fig:notation}.

\begin{table}[h]
\centering
\begin{tabular}{l|l}
\toprule
\multicolumn{2}{l}{Protein \& Motif}\\
  $\pP$ & A protein with both structure and sequence \\
  $\pA$ & Protein amino acid sequence. $\mathcal{A} = \{\mathcal{C}^{n}, \mathcal{D}\}$ \\
  $\pX$ & Protein 3D structure \\
  $a_i \in \mathcal{C}^{20}$ & The $i$-th amino acid type \\
  $\bx_i \in \mathbb{R}^3$ & The $i$-th C-$\alpha$ residue backbone coordinates in 3D \\
  $\mathcal{D}$ & Sequence position. The index of amino acids in the sequence \\
  $\pM_\pP$ & Scaffolded motif in a protein \\
  $\pM_{\rm X}$ & Internal structure of motif without scaffolding\\
  $\pM_\pA$ & Internal sequence of motif without scaffolding \\
  $\pA_\pM$ & Scaffolded motif position in the protein sequence \\
  $\pX_\pM$ & Scaffolded motif position in the protein structure \\

  $\pS_\pP$ & Scaffolding in a protein \\
  $\pS_{\rm X}$ & Internal structure of scaffolding without motif\\
  $\pS_\pA$ & Internal sequence of scaffolding without motif \\
  $\pA_\pS$ & Scaffolding position in the protein sequence with motif \\
  $\pX_\pS$ & Scaffolding position in the protein structure with motif \\
\multicolumn{2}{l}{Parameterization}\\
  $\pT \in \mathbb{R}^{4\times4}$ & Residue portions (transformation). Orientation preserving rigid transformation (\textit{frame}) \\
  $\pR_i \in \mathbb{R}^{3\times 3}$ & Rotation \\
  $\pX_i \in \mathbb{R}^3$ & Translation \\
\bottomrule
\end{tabular}
\caption{Notations for FADiff}
\label{tab:notation}
\end{table}

\section{Method details}

\subsection{Noise on rotation}
\label{app:noise}
Given the rotation $\Ri$ and $\Ro$, we can calculate the transformation from $\Ro$ to $\Ri$ as:
\begin{equation}
    \rn' = \invRo \Ri .
\end{equation}
With Eq.~\eqref{eq:rotationri}, the equation above is further derived as:
\begin{equation}
    \rn' = \invRo \left[\Delta\pR_\pM, \Delta\pR_\pS\right] \Ro 
\end{equation}
The transformation is consistent with the diffusion process of FrameDiff.
With Eq.~\eqref{eq:delta}, we have:
\begin{align*}
    \rn' &= \invRo \left[(\overline{\Ro\rn\invRo})_\pM, (\Ro\rn\invRo)_\pS\right] \Ro \\
        &=  \left[\left(\invRo\overline{\Ro\rn\invRo}\Ro\right)_\pM, (\invRo \Ro\rn\invRo)\Ro_\pS \right] \\
        &= \left[\left(\invRo\overline{\Ro\rn\invRo}\Ro\right)_\pM, (\rn)_\pS \right] 
\end{align*}

\subsection{Translation update}
\label{app:Translation}
Eq.~\eqref{eq:xl} is derived as follows:
Since the model gives an update under the coordinate system of the local frame. We first get the update of translation under the fixed coordinate system through Eq.~\eqref{equ:transform} as:
\begin{equation*}
    \label{eq:exptrans}
    \tnh^{world} = \Rlo \tnh
\end{equation*}
There are two parts of translation updates: (1) the model expected translation update $\tnh^{world}$ (Eq.~\eqref{eq:exptrans}), and (2) the rigid motif rotation caused translation $\Delta\hat{\pX}_\pM$ (Eq.~\eqref{eq:rotationcause}). We believe the model's expected translation update is translating the residues to rational positions. In contrast, the translation from the rigid anchor motif rotation interferes with the movement to the rational position. Therefore, we remove the translation update caused by rotation, then average them on the motif residues as:
\begin{align}
    \label{eq:avgrot}
    \tnh^{update} = [(\overline{\tnh^{world}- \Delta\hat{\pX}_\pM})_\pM, (\tnh^{world})_\pS]
\end{align}
Finally, we add the translation update to $\Xlo$ to get $\Xli$ as:
\begin{align}
    \Xli & = \tnh^{update} + \Xlo \notag \\
         & \stackrel{Eq.~\eqref{eq:avgrot}}{=}  [(\overline{\tnh^{world}- \Delta\hat{\pX}_\pM})_\pM, (\pX_\delta^{world})_\pS] + \Xlo \notag\\
         & \stackrel{Eq.~\eqref{eq:exptrans}}{=} \left[\overline{(\Rlo\tnh)_\pM-\Delta\hat{\pX}_\pM}+\Delta\hat{\pX}_\pM, (\Ro\tnh)_\pS\right] \notag + \Xlo .
\end{align}

\subsection{Translation for score calculation}
\label{app:score_trans}
Given the predicted rotation matrix $\hat{\pR}^{(0)}$ from the model, the noised rotation matrix $\Ri$, and the noised translation $\Xi$, we can derive the translation caused by the rigid anchor motif as follows:
The rotation from $t=0$ to $t=t$ is:
\begin{equation}
    \label{ot}
    \pR_{0\rightarrow t} = (\hat{\pR}^{(0)})^{-1} \Ri
\end{equation}
Then the translation caused by the rotation is:
\begin{align*}
    \Delta \pxm' & = \pR_{0\rightarrow t}^{-1}(\Xi - \pX_v) + \pX_v - \Xi \\
                &\stackrel{Eq.~\eqref{ot}}{=} ((\hat{\pR}^{(0)})^{-1} \Ri)^{-1} (\Xi - \pX_v) + \pX_v - \Xi \\
                & = (\Ri)^{-1} \hat{\pR}^{(0)} (\Xi - \pX_v) + \pX_v - \Xi
\end{align*}

\subsection{Relationship between FADiff and other score-based diffusion models}
\label{app:consistent}
In this work, we adopt the FrameDiff \citep{se3} as the pipeline for the diffusion process. Without loss of generalization, any other score-based diffusion models can be used here. 

We explain that FADiff is consistent with FrameDiff below
First, we review the FrameDiff.
In the forward diffusion process, the noise is added to translation and rotation as:
\begin{align}
    \Ri &= \Roo \rn(t) \notag \\
    \Xi &= \Xoo + \tn(t), 
\end{align}
where the score $s$ for translation and rotation can be calculated as:
\begin{align}
    s_\br^{Frame} &= \operatorname{score}_\br(\rn(t), t) \notag \\
    s_\bx^{Frame} &= \operatorname{score}_\bx(\Xi, \Xoo, t).
\end{align}

The score network gives the predicted denoised rotation and translation $\Rooh$ and $\Xooh$, with which the score can be calculated as:
\begin{align*}
    s_\br'^{Frame} &= \operatorname{score}_\br((\Rooh)^{-1}\Ri, t) \notag \\
    s_\bx'^{Frame} &= \operatorname{score}_\bx(\Xi, \Xooh, t).
\end{align*}
Since all the layers for the update are the same, we take the last layer for example, and then $\Roh$ can be derived as:
\begin{align}
   \Rli &= \Rlo\rnh \\
   \Rooh &= \Ri\rnh
\end{align}
Then the score can be calculated as:
\begin{align}
    s_\br'^{Frame} &= \operatorname{score}_\br(\rnh^{-1}, t) \notag \\
    s_\bx'^{Frame} &= \operatorname{score}_\bx(\Xi, \Xooh, t).
\end{align}

FrameDiff seeks to match $s_\br$ and $s_\bx$ to $s_\br'$ and $s_\bx'$. The object to optimize is:
\begin{align}
    \min \quad & \Vert s_\br - s_\br' \Vert^2 + \Vert s_\bx - s_\bx' \Vert^2 \notag \\
    \label{eq:object}
    \min \quad \Vert \operatorname{score}_\br(\rn(t), t) - \operatorname{score}_\br(\rnh^{-1} &, t) \Vert^2 + \Vert \operatorname{score}_\bx(\Xi, \Xoo, t) - \operatorname{score}_\bx(\Xi, \Xooh, t) \Vert^2 
\end{align}
For an ideal FrameDiff, $\rnh = \rn^{-1}$ and $\hat{\pX}^{(0)}=\pX^{(0)}$. 
To keep consistent with FrameDiff, the score network should predict $\rnh = \rn^{-1}$ and the whole model should predict the $\hat{\pX}^{(0)}=\pX^{(0)}$. 

Generally, the model should predict the noise added to the rotation and translation in the forward diffusion process. In FADiff, we get consistency by making the model predict the noise actually added to the original data, \ie, the $\rn'$ and $\tn'$. The whole process is listed below.
In the forward process, the rotation and translation noise for each residue of motifs are:
\begin{align}
    \rn' &= \left[\left(\invRoo\overline{\Roo\rn\invRoo}\Roo\right)_\pM,(\rn)_\pS \right], \notag\\
    \tn' &= \left[\overline{\tn}_\pM, \tn_\pS \right].
\end{align}
Then the noised data is derived as:
\begin{align}
    \label{eq:xt}
    \Ri &= \left[(\overline{\Roo\rn\invRoo}\Roo)_\pM,(\Roo\rn)_\pS \right], \notag\\
    \Xi &= \left[\left( \Xoo + \Delta \pX_\pM +\overline{\tn}\right)_\pM, \left(\Xoo + \tn\right)_\pS \right],
\end{align}
where $\Delta \pX_\pM$ is the translation caused by the rotation of rigid anchor motif as Eq.~\eqref{eq:forward} and Eq.~\eqref{eq:noise1r}:
\begin{equation*}
    \Delta \pX_\pM = (\overline{\Roo\rn\invRoo})_\pM (\Xoo_\pM - \pX_v) + \pX_v - \Xoo_\pM
\end{equation*}

The score is calculated as:
\begin{align}
    \label{eq:inputscore}
    s_\br^{FA} &= \operatorname{score}_\br(\left[\left(\invRoo\overline{\Roo\rn\invRoo}\Roo\right)_\pM,(\rn)_\pS \right], t), \notag \\
    s_\bx^{FA} &= \operatorname{score}_\bx(\left[\left(\Xoo + \overline{\tn}\right)_\pM, \left(\Xoo + \tn\right)_\pS \right], \Xoo, t),
\end{align}
without the translation caused by the rotation.
The score network gives the predicted denoised rotation and translation $\Roh$ and $\Xoh$.

The predicted rotation and translation is the updated frame from the last layer as Eq.~\eqref{rq:lastlayer} and Eq.~\eqref{eq:xl}:
\begin{align}
   \Rli &= \left[ (\overline{\Rlo\rnh\invRlo})_\pM, (\Rlo\rnh\invRlo)_\pS \right]\Rlo \notag \\
    \Xli &= \left[\overline{(\Rlo\tnh)_\pM-\Delta\hat{\pX}_\pM}+\Delta\hat{\pX}_\pM, (\Rlo\tnh)_\pS\right] + \Xlo
\end{align}
Since all the layers for the update are the same, we take the last layer for example, and then the equation above is derived as:
\begin{align}
    \label{eq:31}
    \Rooh &= \left[\left(\overline{\Ri\rnh\invRi}\Ri\right)_\pM, (\Ri\rnh)_\pS \right]\\
    \Xooh &= \left[\overline{(\Ri\tnh)_\pM-\Delta\hat{\pX}_\pM}+\Delta\hat{\pX}_\pM, (\Ri\tnh)_\pS\right] + \Xi
\end{align}
With Eq.~\eqref{eq:transupdaterota}, the translation of each residue caused by the rotation is:
\begin{equation}
    \label{eq:33}
    \Delta \hat{\pX}_\pM = (\Ri)^{-1} \hat{\pR}^{(0)} (\Xi - \pX_v) + \pX_v - \Xi.
\end{equation}

Finally, the predicted scores from the model are :
\begin{align}
    \label{eq:score_fadiff}
    s_\br'^{FA} &= \operatorname{score}_\br(\left[\left(  \left(\overline{\Ri\rnh\invRi}\Ri\right)^{-1}_\pM  \Ri\right)_\pM,(((\Ri\rnh)_\pS)^{-1}\Ri)_\pS \right], t), \notag \\
              & = \operatorname{score}_\br(\left[\left(  \left(\overline{\Ri\rnh\invRi}\Ri\right)_\pM^{-1}  \Ri\right)_\pM,(\rnh^{-1})_\pS \right], t), 
              \\
    \label{eq:last}
    s_\bx'^{FA} &= \operatorname{score}_\bx(\left[\left(\Xoo + \overline{\tn}\right)_\pM, \left(\Xoo + \tn\right)_\pS \right], \Xooh-[\Delta \hat{\pX}_\pM, 0_\pS] , t) \notag \\
              &= \operatorname{score}_\bx(\left[\left(\Xoo + \overline{\tn}\right)_\pM, \left(\Xoo + \tn\right)_\pS \right], [(\Xooh-\Delta \hat{\pX})_\pM, \Xooh_\pS] , t)
\end{align}
We can explain the derivation in two parts. The scaffolding part is consistent with FrameDiff obviously with Eq.~\eqref{eq:score_fadiff} and Eq.~\eqref{eq:last}, which is the same as FrameDiff. For the motif part, substituting Eq.~\eqref{eq:xt}, Eq.~\eqref{eq:31} and Eq.~\eqref{eq:33} into Eq.~\eqref{eq:score_fadiff} and Eq.~\eqref{eq:last} yields the same form as Eq.~\eqref{eq:inputscore}. When we relax the equation with the average to be themselves, and further substitute $\hat{\pX}^{(0)}=\pX^{(0)}$ and $\rnh = \rn^{-1}$ into the equation, the equation gets the same as Eq.~\eqref{eq:inputscore}. 

\subsection{Sampling}
\label{app:sample}
\subsubsection{TSP for sequence construction}
In sampling, we randomly sample residues of the scaffold with a Gaussian distribution to decide their translation and rotation. The motifs are put at the origin of the fixed coordinate. Then we construct a distance map by calculating the distance between the atoms C and N of two residues, which should be the length of the peptide bond in a naturally existing protein. Besides, to maintain the motif structure, we change the distance between two residues connected in the motif to \textit{0} and the distance from the scaffold residues to the connected motif residues to be infinite. Finally, with a greedy algorithm to find the shortest chain like the TSP, a sequence is obtained.
\subsubsection{Random connection for sequence construction}
\label{app:randomsample}
In sampling, we randomly sample residues of the scaffold with a Gaussian distribution to decide their translation and rotation. The motifs are put at the origin of the fixed coordinate. 
Then we construct a distance map where we set the distance between two residues connected in the motif to \textit{0} and the distance from the scaffold residues to the connected motif residues to be infinite to maintain the motif structure. And the distances between other residues are given randomly. Finally, with a greedy algorithm to find the shortest chain like the TSP, a sequence is obtained.

\section{Training details}
\subsection{Hyper-parameters}
\label{app:hyper-param}
We follow the FrameDiff \citep{se3} for all the parameters except the coordinate scale.
We train FADiff for 90,000 steps with a coordinate scale of 0.1 and 0.02 based on the pre-trained VFN-Diff \citep{vfn}. Here the coordinate scale $c$ is used in the adding noise stage: 
$$\bx^t = c \cdot  (\bx^{(t-1)} \cdot c + \bx_\delta),$$
where $x_\delta$ is the Gaussian noise.
The coordinate scale of 0.1 follows \citet{se3}; however, the model trained with the coordinate scale of 0.02 works much better since the coordinate scale is much larger than a single amino acid. 

\subsection{Hardware}
We train FAdiff for 90,000 steps in around 20 hours. All our experiments are conducted on a computing cluster with 8 GPUs of NVIDIA
GeForce
RTX 4090 24GB and CPUs of AMD
EPYC
7763 64-Core of 3.52GHz. 
All the inferences are conducted on a single GPU of NVIDIA
GeForce
RTX 4090 24GB.

\subsection{Training loss}
\label{app:loss}
Since the method of FADiff is general, any other score-based diffusion model can be adopted as our backbone. All the training loss weights and other settings remain the same except the coordinate scale as explained in the Appendix \ref{app:hyper-param}

\section{Experiment}
\subsection{Dataset Detail}
\subsubsection{Distribution of sequence length of the VM dataset}
\label{app:dataset}
\begin{figure}[ht!]
    \centering
    \includegraphics[width=0.52\linewidth]{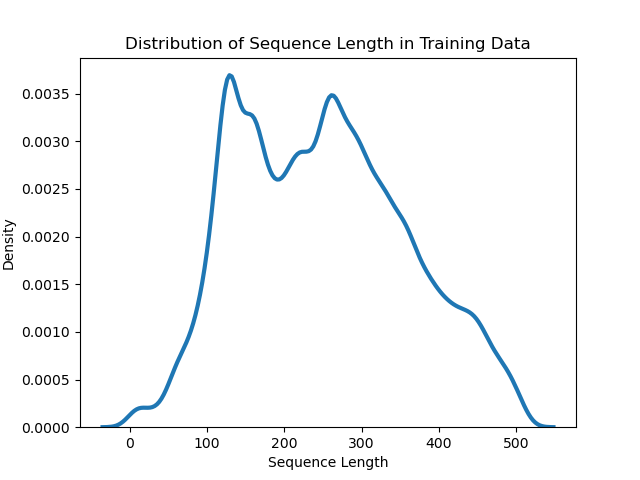}
    \caption{The distribution of sequence length for the VM dataset.}
    \label{fig:traing}
\end{figure}

The number of proteins in the VM dataset varies with the sequence length as shown in Fig.~\ref{fig:traing}.
\subsubsection{Motif dataset}
The motif dataset is extracted from PROSITE, a database maintained by Swiss Institute of Bioinformatics (SIB), which contains 1942 documentation entries, 1311 patterns, and 1400 ProRules (dated January 24, 2024). It contains patterns, profiles, and rules for recognizing specific motifs in protein sequences. 

The dataset consists of 16,251 motif fragments, based on their representation in the Protein Data Bank (PDB), which involved aligning protein sequences and atom coordinates with known motifs.

\subsection{Evaluation Metrics}
\label{app:evaluation}
\begin{figure}[ht!]
    \centering
    \includegraphics[width=0.41\linewidth]{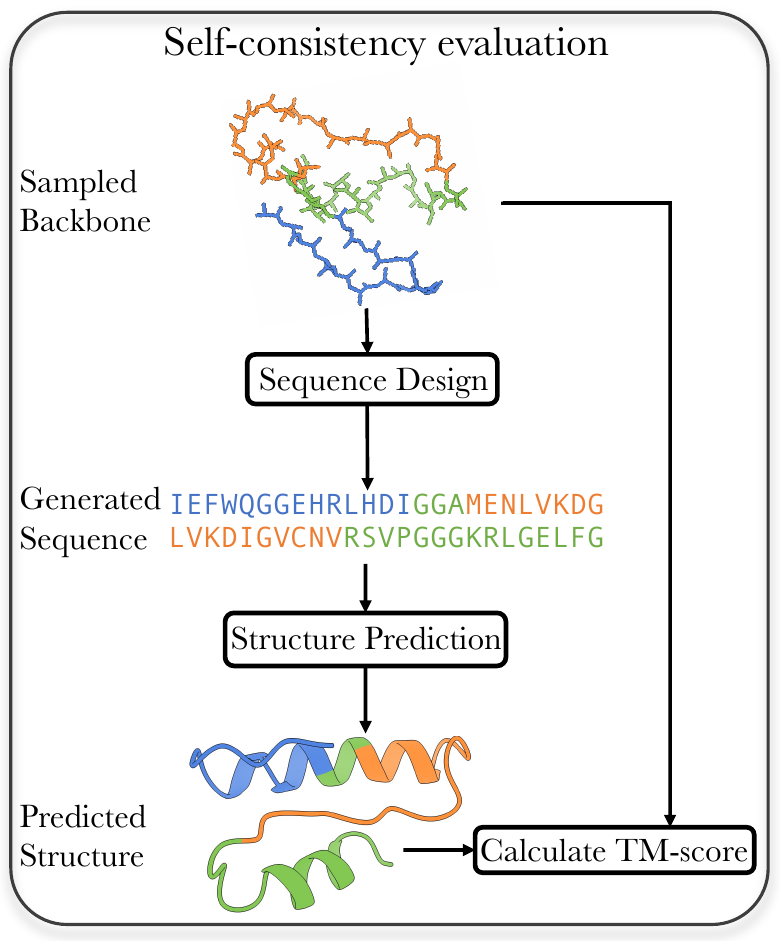}
    \caption{For self-consistency evaluation, we utilize a pre-trained fixed-backbone sequence-design model, namely ProteinMPNN, to design the scaffold sequence from the generated protein structure. Then we put the designed sequence to ESMFold to obtain the full protein structure. Finally, we calculate the TM-score between the predicted structure and the original backbone structure. In the figure, the \textcolor{orange}{orange}, \textcolor{blue}{blue}, and \textcolor{green}{green} colors indicate the motif \texttt{2BGS}, \texttt{1G79}, and the scaffold.}
    \label{fig:evaluation}
\end{figure}

Following \citet{smcdiff}, we calculate the \textbf{scTM} of one generated structure as follows: (1) we utilize the ProteinMPNN \citep{proteinmpnn} to design the amino acid sequence. (2) The designed sequence from the ProteinMPNN is input into the ESMFold \citep{esmfold} to get the structure. (3) The TM-score between the structures from the ESMFold and generated from our model is calculated as the scTM. The workflow is illustrated in Fig.~\ref{fig:evaluation}.

We calculate the \textit{in silico} Success Rate following \citet{smcdiff} as follows: (1) for each group of motifs, we generate 5 samples for each length. (2) each generated sample is input into the ProteinMPNN to design the sequence. (3) each sequence is input into the ESMfold 8 times to get the folding protein structures. (4) We calculate the TM score between our generation and the 8 folding results from ESMFold and take the highest TM score as scTM from this generation. One generation is successful if the $\rm scTM>0.5$

We have also calculated the \textbf{diversity} and \textbf{pdbTM} for each generation. \textbf{Diversity} is the ratio of unique clusters in the number of generated samples where the clusters are produced by MaxCluster \citep{div} following previous works. 
pdbTM indicates the novelty of generated protein structures. Each generated protein structure is compared with the structures in PDB~\cite{pdb} to get the TM-score between two structures as pdbTM score, one generation is novel if the $\rm pdbTM<0.7$.

\subsection{Experimental Results}
\label{app:Results}
\subsubsection{Evaluation on VM Dataset}
\label{app:tmscore_eval}
\begin{figure}[ht!]
    \centering
    \includegraphics[width=0.65\linewidth]{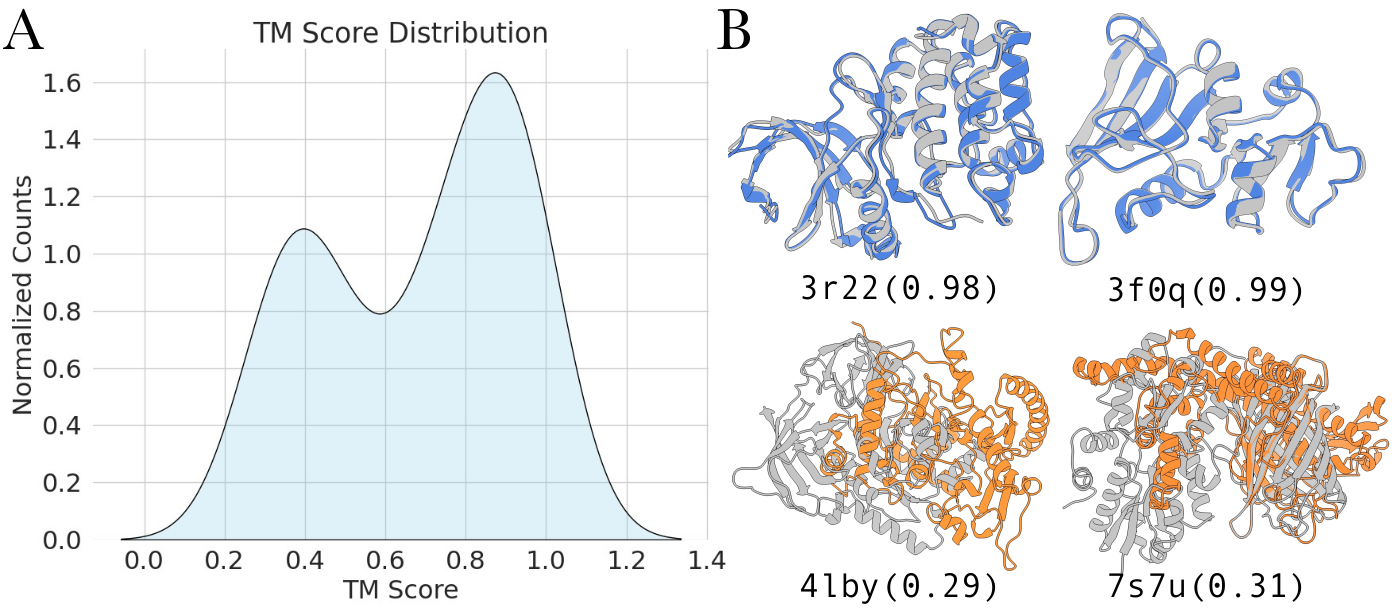}
    \caption{A. TM-score distribution on evaluation dataset of VM dataset. B. The generated structure with a high tm-score closely matches the native structure, while a low tm-score indicates the generated structure differs from the native structure (grey). The four-digit number below each structure indicates the protein identifier and the number in the brackets denotes the TM score.  }
    \label{fig:tmscore}
\end{figure}

The distribution of TM-scores on the VM dataset is shown in Fig.~\ref{fig:tmscore}A. Hign TM-score indicates the generated protein structure is similar to the native structures where the desired motifs are located.

\subsubsection{Evaluation on MS Dataset}
\label{app:diversity_exp}
The diversity of all the generations is 1, so we just mention it in the appendix. The pdbTM which indicates the Novelty of generation for each generation, is shown in Table \ref{tab:pdbtm}. Following previous works \citet{se3,vfn,rfdiff}, one generated structure is novel if the pdbTM$<0.7$, here we show the ratio of novel protein structures in the generation comparing with FrameDiff and VFNDiff.
\begin{table}[ht!]
    \centering
    \begin{tabular}{cccccccc}
        \toprule
        \multicolumn{2}{c}{Method}& \multicolumn{4}{c}{FADiff} & \multirow{2}{*}{VFN-Diff} & \multirow{2}{*}{FrameDiff}\\
        \multicolumn{2}{c}{\# Motif} & 2 & 3 & 4 & 5 & &  \\ 
        \midrule 
        \multirow{2}{*}{pdbTM$_{\rm <0.7}$} & overall & 84.60\% & 89.23\% & 90.33\% & 93.75\% & 41.67\% & 54.67\% \\ 
         & designable & 82.94\% & 85.66\% & 86.88\% & 89.91\% & 1.67\% & 1.33\% \\

        \bottomrule
    \end{tabular}
    \caption{pdbTM for each generation. The numbers indicate the ratio of novel protein structures in the generated samples.}
    \label{tab:pdbtm}
\end{table}

\subsubsection{Ablation Study}
\label{app:Ablation}
The whole results of the ablation study on all the lengths are shown in Table \ref{tab:ablation_full}.

\begin{table}[ht!]
    \centering
    \begin{tabular}{clccccccc}
        \toprule
        Method& 160 & 210 & 260 & 310 & 360 & 410 & Avg\\
        \midrule 
        FADiff  &76.67 & 71.67 & 81.67 & 65.00 & 86.67 & 56.67 & 73.05\\
        Random  & 49.33 & 60.00 & 60.00 & 54.67 & 82.67 & 57.33 & 60.67\\ 
        $/$2  &  69.23 & 76.92 & 76.92 & 60.00 & 83.08 & 69.23 & 72.56\\
        $\times$2 & 70.00 & 70.00 & 75.00 & 78.33 & 83.33 & 56.67 & 72.22\\ 
        \bottomrule
    \end{tabular}
    \caption{\textit{In silico} success rate (SR\%) for different lengths, with/without TSP, and with different translation noise scales in sampling. TSP and Random indicate the sequence construction method. $/$2 and $\times$2 indicate the noise scale on translation in sampling. The Avg indicates the Average success rate for one method.}
    \label{tab:ablation_full}
\end{table}

\begin{figure}[ht!]
    \minipage{0.49\textwidth}
    \centering
    \includegraphics[width=0.5\linewidth]{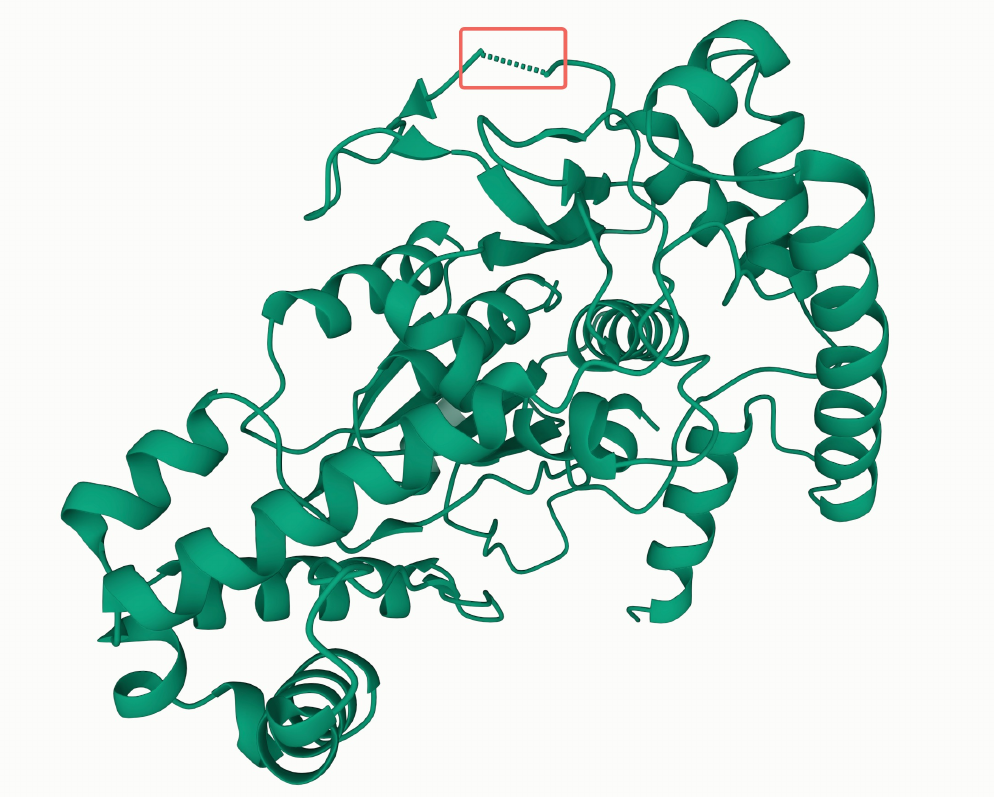}
    \caption{Failed case 1}
    \label{fig:fail1}
    \endminipage\hfill
    \minipage{0.49\textwidth}
    \centering
    \includegraphics[width=0.5\linewidth]{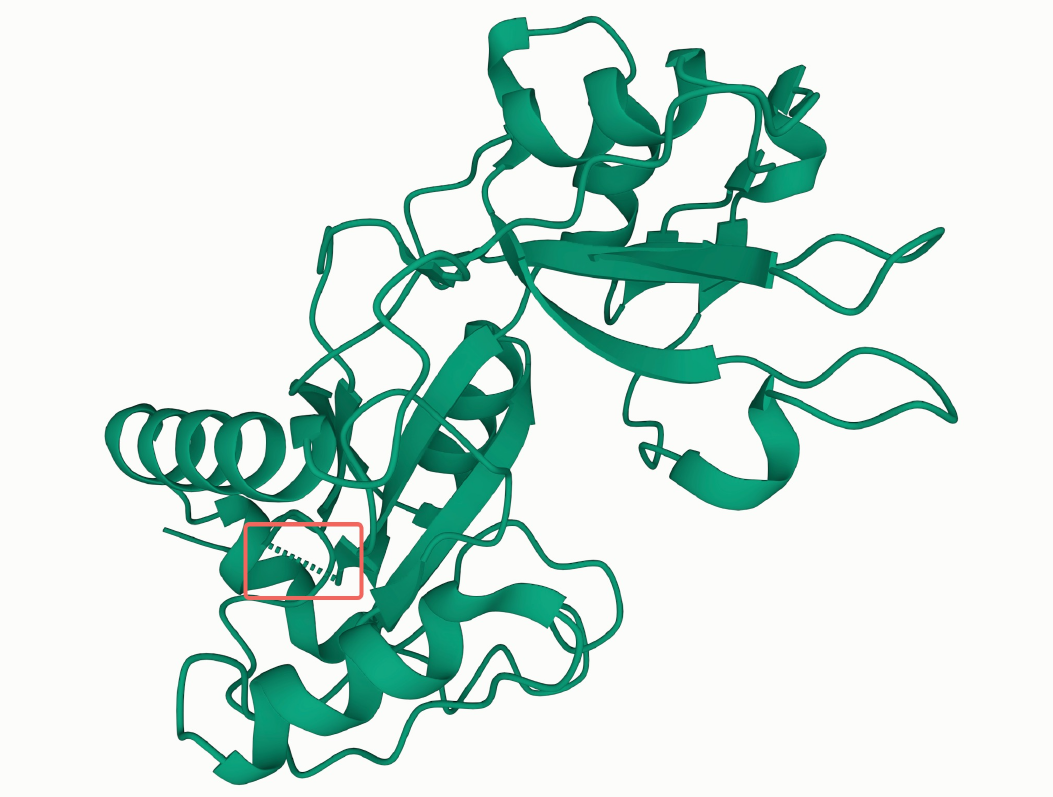}
    \caption{Failed case 2}
    \label{fig:fail2}
    \endminipage
\end{figure}

\paragraph{Noise scale on translation in training}
The failed cases from the FADiff trained without increasing the noise scale on translation are shown in Fig.~\ref{fig:fail1} and Fig.~\ref{fig:fail2}.

\begin{figure}[ht!]
    \centering
    \includegraphics[width=0.75\linewidth]{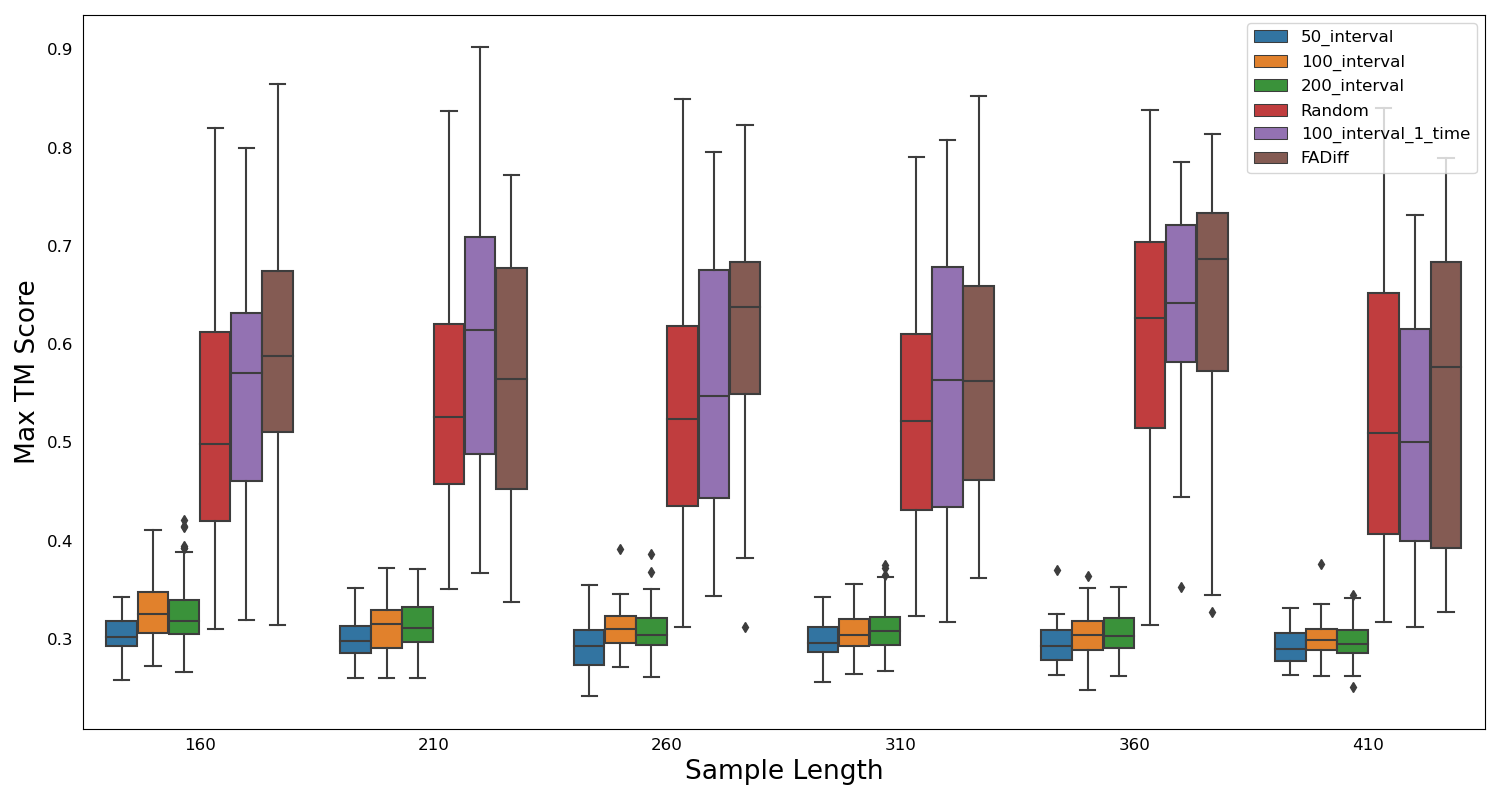}
    \caption{50\_interval, 100\_interval, and 200\_interval indicate that we perform TSP every 50, 100, and 200 steps in sampling. Random indicates TSP is not performed. 100\_interval\_1\_time and FADiff indicate that we only perform TSP in the first 100 steps two times or multiple times (every step) in the first 100 steps.}
    \label{fig:tsp_compare}
\end{figure}

\paragraph{TSP for sequence construction}
\label{app:tsp}
TSP is performed in the first 100 steps while random is performed at the first step only. Here, we show the results of performing TSP with different settings in Fig.~\ref{fig:tsp_compare} and Table \ref{tab:tsp_compare}. If we conduct TSP throughout the entire sampling process, this will cause the model to be unable to converge because the position of amino acids on the sequence is constantly changing. However, if we conduct TSP in the first 100 steps, this will steer the motifs to the appropriate position because the noise is gradually decreasing during the first 100 steps. Besides, conducting TSP only in the first 100 steps will not cause the model to fail to converge. Performing TSP multiple times results in a better scTM than only performing TSP once in the first 100 steps.

\begin{table}[ht!]
    \centering
    \begin{tabular}{clccccccc}
        \toprule
        Method& 160 & 210 & 260 & 310 & 360 & 410 \\
        \midrule 
        Random & 49.33 & 60.00 & 60.00 & 54.67 & 82.67 & 57.33 \\
        100\_interval\_1\_time & 67.50 & 67.50 & 60.00 & 62.50 & 90.00 & 50.00 \\
        \cellcolor{lm_purple_low}FADiff  & \cellcolor{lm_purple_low}76.67 & \cellcolor{lm_purple_low}71.67 & \cellcolor{lm_purple_low}81.67 & \cellcolor{lm_purple_low}65.00 & \cellcolor{lm_purple_low}86.67 & \cellcolor{lm_purple_low}56.67 \\
        \bottomrule
    \end{tabular}
    \caption{\textit{In silico} success rate (SR\%) for different TSP settings.}
    \label{tab:tsp_compare}
\end{table}
\subsubsection{Comparison}
\begin{figure}[ht!]
    \centering
    \includegraphics[width=0.7\linewidth]{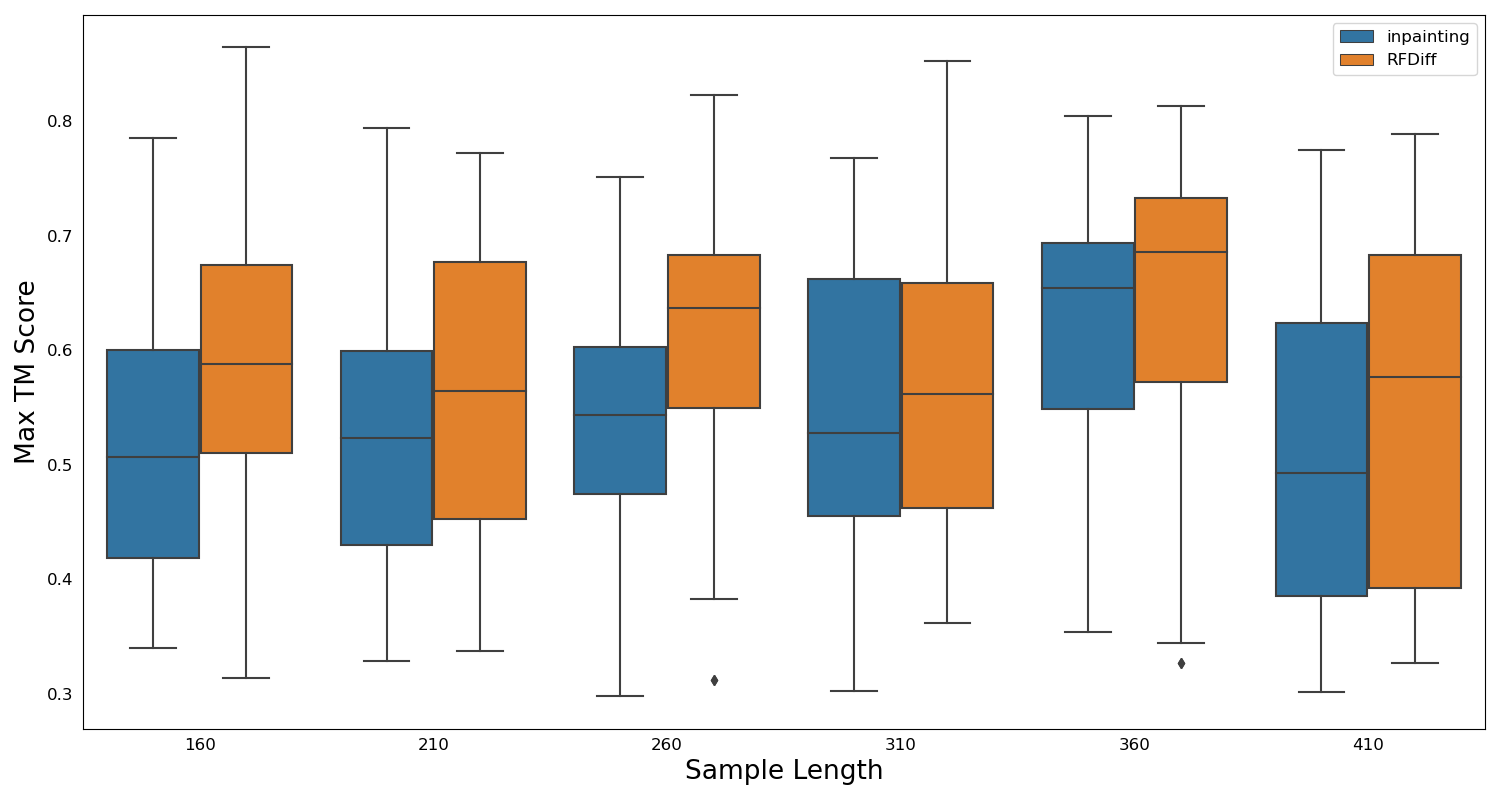}
    \caption{TM-score of inpainting methods and FADiff on generation different lengths of scaffoldings.}
    \label{fig:tsp_compare_app}
\end{figure}

We also compare FADiff with inpainting and conditional generation methods. FADiff outperforms inpainting consistently on the SCTM score as shown in Fig.~\ref{fig:tsp_compare_app}.

\end{document}